\documentclass[journal]{IEEEtran}
 
\usepackage{cite}
\usepackage{graphicx}
\usepackage{amsmath,amssymb,amsfonts}
\usepackage{algorithm}
\usepackage{algorithmicx}
\usepackage{algpseudocode}
\usepackage{listings}
\usepackage{multirow}
\usepackage{array}
 
\ifCLASSINFOpdf
\else
\fi
 
\usepackage{amsmath,amssymb,amsfonts}
\usepackage{algorithm}
\usepackage{algorithmicx}
\usepackage{algpseudocode}
\usepackage{listings}

\begin{document}

\title{Physics-Informed Graph Neural Jump ODEs for Cascading Failure Prediction in Power Grids}
 
\author{Birva~Sevak, Shrenik Jadhav, and Van-Hai~Bui \IEEEmembership{Senior Member, IEEE}%
\thanks{Birva Sevak and Shrenik Jadhav are with the Department of Computer and Information Science, University of Michigan--Dearborn, Michigan, USA.}%
\thanks{Van-Hai Bui is with the Department of Electrical and Computer Engineering, University of Michigan--Dearborn, Michigan, USA. 
 
Corresponding author: Van-Hai Bui (e-mail: vhbui@umich.edu).}%
}%
 
\maketitle

\begin{abstract}
Cascading failures in power grids pose severe risks to infrastructure reliability, yet real-time prediction of their progression remains an open challenge. Physics-based simulators require minutes to hours per scenario, while existing graph neural network approaches treat cascading failures as static classification tasks, ignoring temporal evolution and physical laws. This paper proposes Physics-Informed Graph Neural Jump ODEs (PI-GN-JODE), combining an edge-conditioned graph neural network encoder, a Neural ODE for continuous power redistribution, a jump process handler for discrete relay trips, and Kirchhoff-based physics regularization. The model simultaneously predicts edge and node failure probabilities, severity classification, and demand not served, while an autoregressive extension enables round-by-round temporal cascade prediction. Evaluated on the IEEE 24-bus and 118-bus systems with 20,000 scenarios each, PI-GN-JODE achieves a Precision--Recall Area Under the Curve of 0.991 for edge failure detection, 0.973 for node failure detection, and a coefficient of determination of 0.951 for demand-not-served regression on the 118-bus system, outperforming a standard graph convolutional network baseline (0.948, 0.925, and 0.912, respectively). Ablation studies reveal that the four components function synergistically, with the physics-informed loss alone contributing +9.2 points to demand-not-served regression. Performance improves when scaling to larger grids, and the architecture achieves the highest balanced accuracy (0.996) on the PowerGraph benchmark using data from a different simulation framework.
\end{abstract}
 

\IEEEpeerreviewmaketitle

\section{Introduction}
 
\IEEEPARstart{C}{ascading} failures in power grids remain one of the most consequential threats to modern infrastructure. A single initiating event, such as a transmission line tripping due to thermal overload or a generator outage, can trigger a chain reaction of successive component failures that propagates across the network within minutes. The resulting blackouts can affect tens of millions of people and impose billions of dollars in economic losses. The August 2003 blackout in the northeastern United States and Canada left approximately 50 million people without power for up to four days after a sequence of line trips escalated into the largest blackout in North American history~\cite{uscanadarpt2004}. More recently, Winter Storm Uri in February 2021 caused the loss of over 61,000~MW of generation capacity across 1,045 units in the south-central United States, leaving roughly 4.5 million customers in Texas without electricity for days~\cite{fercnerc2021}. Statistical analyses of historical blackout data have consistently shown that the size distribution of such events follows a heavy-tailed power law, meaning that large-scale cascading failures occur far more frequently than normal distributions would predict~\cite{dobson2007complex}. This persistent vulnerability, combined with the growing complexity introduced by renewable energy integration and distributed generation, makes it increasingly urgent to develop predictive tools that can anticipate cascade progression and support proactive intervention by grid operators.
 
Traditional approaches to cascading failure analysis rely on detailed physical simulation, including the OPA model~\cite{carreras2004complex}, the Manchester model~\cite{nedic2006criticality}, and DCSIMSEP~\cite{eppstein2012random}, among others surveyed by the IEEE PES CAMS Task Force~\cite{vaiman2012risk}. Although these approaches faithfully capture the physics of cascade propagation, their computational cost grows rapidly with system size. Running thousands of Monte Carlo simulations on large-scale grids can require hours or days of computation, making them impractical for real-time decision support during evolving grid emergencies.
 
This computational bottleneck has motivated machine learning approaches, from early SVM-based methods~\cite{gupta2015svm} to graph neural networks (GNNs) that respect the underlying network topology~\cite{donon2020neural, chadaga2024cascade}. The PowerGraph benchmark~\cite{varbella2024powergraph} provides standardized datasets for systematic GNN evaluation on power grids, and our prior work~\cite{jadhav2025enhancing} demonstrated topology-aware gated message passing for AC power flow estimation. Despite these advances, most existing GNN approaches treat cascading failure prediction as a static, one-shot classification task without modeling the temporal progression of the cascade, and purely data-driven models often lack physical consistency.
 
A key insight motivating the present work is that cascading failures are inherently dynamic processes: between successive failures, power flows redistribute continuously according to Kirchhoff's laws, while at discrete moments, relay activations cause abrupt topological changes. Neural ODEs~\cite{chen2018neuralode} and their graph extensions~\cite{poli2019graph, xhonneux2020cgnn} provide frameworks for modeling continuous-time dynamics on graph-structured data, while neural jump SDEs~\cite{jia2019neural} handle discrete discontinuities through state-dependent jump processes.
 
Physics-informed machine learning~\cite{raissi2019physics, karniadakis2021piml} offers a complementary strategy by embedding domain knowledge into the learning process. In power systems, PINNs have been applied to transient stability~\cite{misyris2020pinns} and power flow analysis~\cite{hu2021physics, huang2023review}. For cascading failure prediction, enforcing physical constraints such as power balance at buses ensures that predictions correspond to physically realizable grid states.
 
The sequential nature of cascade propagation has been established by Sch\"{a}fer et al.~\cite{schafer2018dynamically}, who showed that nonlinear transient dynamics can trigger cascade modes invisible to static analysis, and Hines et al.~\cite{hines2017cascading}, who demonstrated that the effective propagation topology of cascading outages differs from the physical grid topology. These findings motivate models that predict the round-by-round progression of failures. From an architectural perspective, standard GCNs~\cite{kipf2017semi} cannot represent heterogeneous line properties; edge-conditioned convolution~\cite{simonovsky2017dynamic} and neural message passing~\cite{gilmer2017neural, battaglia2018relational} address this by incorporating branch parameters directly into the message-passing computation.
 
In this paper, we propose Physics-Informed Graph Neural Jump ODEs (PI-GN-JODE), which unifies these elements into a single framework for cascading failure prediction. The model combines an edge-conditioned GNN encoder, a Neural ODE for continuous dynamics, a jump handler for discrete relay trips, and Kirchhoff-based physics regularization, simultaneously predicting edge failures, node failures, severity, and demand not served. We validate on the IEEE 24-bus~\cite{ieee24rts1979} and 118-bus systems using pandapower~\cite{thurner2018pandapower}, benchmark against the PowerGraph dataset~\cite{varbella2024powergraph}, and demonstrate multi-round cascade prediction via autoregressive unrolling with scheduled sampling~\cite{bengio2015scheduled}. To our knowledge, this is the first framework combining GNNs, Neural ODEs, jump processes, and physics constraints for temporal cascade prediction.
 
The main contributions of this paper are as follows:
\begin{enumerate}
    \item We introduce PI-GN-JODE, a unified architecture combining edge-conditioned GNNs, Neural ODE continuous dynamics, discrete jump handlers, and Kirchhoff-based physics constraints for multi-task cascading failure prediction.
    \item We demonstrate through ablation studies that each component makes a synergistic contribution, with the physics-informed loss providing the largest single improvement to DNS regression (+9.2 points in $R^2$).
    \item We validate on IEEE 24-bus and 118-bus systems (Edge PR-AUC 0.991, Node PR-AUC 0.973, DNS $R^2$ 0.951 on 118-bus) and extend to multi-round temporal cascade prediction via autoregressive unrolling with scheduled sampling.
 
\end{enumerate}
 
Section~\ref{sec:methodology} presents the methodology, Section~\ref{sec:results} reports experimental results, and Section~\ref{sec:conclusion} concludes with future directions.

\section{Methodology}
\label{sec:methodology}

\subsection{Problem Formulation}
\label{subsec:problem}
 
We represent a power grid as a directed graph $\mathcal{G} = (\mathcal{V}, \mathcal{E})$ where each bus corresponds to a node $v_i \in \mathcal{V}$ and each transmission branch (line or transformer) corresponds to a pair of directed edges $(v_i, v_j), (v_j, v_i) \in \mathcal{E}$. The bidirectional representation enables asymmetric message passing, reflecting the fact that power flows and losses differ by direction. Each node $v_i$ is associated with a feature vector $\mathbf{x}_i \in \mathbb{R}^{d_v}$ derived from the pre-contingency power flow solution, and each directed edge $(v_i, v_j)$ carries a feature vector $\mathbf{e}_{ij} \in \mathbb{R}^{d_e}$ encoding the electrical properties of the corresponding branch.
 
Given the graph $\mathcal{G}$ with an initial contingency specification encoded as binary masks within the features, the model predicts four targets simultaneously:
 
\begin{enumerate}
    \item \textbf{Edge failure probabilities:} $\hat{\mathbf{y}}_{\text{edge}} \in [0,1]^{|\mathcal{E}|}$, the probability that each transmission branch trips during the cascade.
    \item \textbf{Node failure probabilities:} $\hat{\mathbf{y}}_{\text{node}} \in [0,1]^{|\mathcal{V}|}$, the probability that each bus experiences load shedding or isolation.
    \item \textbf{Severity classification:} $\hat{\mathbf{y}}_{\text{sev}} \in \mathbb{R}^{2}$, logits for a binary safe/unsafe classification based on whether demand not served exceeds 5\%.
    \item \textbf{Demand not served (DNS):} $\hat{y}_{\text{dns}} \in [0,1]$, the fraction of total system load lost to the cascade.
\end{enumerate}
 
The framework supports two prediction modes. In \textit{one-shot} mode, a single forward pass maps the initial contingency state to the final cascade outcome. In \textit{multi-round} mode, the model autoregressively predicts the cascade state at each relay operation round, producing a temporal trajectory of failure progression. Both modes share the same base architecture; the multi-round extension unrolls the model's continuous-discrete dynamics across sequential cascade rounds. This formulation goes beyond prior graph-level approaches~\cite{varbella2024powergraph} by providing component-level predictions (which specific lines trip and which buses fail) alongside system-level severity assessments and, in the multi-round setting, the temporal ordering of failures.

\subsection{Cascade Simulation and Dataset Generation}
\label{subsec:data_generation}
 
Training data are generated through Monte Carlo simulation of cascading failures on the IEEE 24-bus Reliability Test System~\cite{ieee24rts1979} and the IEEE 118-bus system using pandapower~\cite{thurner2018pandapower}. For each system, diverse operating points are sampled by randomly scaling the base-case load profile, and a full AC power flow is solved using the Newton-Raphson method to obtain the pre-contingency steady state. Initial N-$k$ contingencies are then selected through a screening procedure that retains only outage scenarios where at least one remaining branch exceeds a minimum effective loading threshold, biasing the dataset toward scenarios with cascade potential. Branches are chosen with probability proportional to their loading, and a leaf-bus bias mechanism preferentially selects lines near low-degree load buses to produce a diverse range of DNS outcomes.
 
Each contingency is propagated through a relay-based thermal overload model in discrete rounds. Branches exceeding 120\% of their thermal rating trip immediately; those between 95--120\% trip if the overload persists for 2--3 rounds. Buses with voltage below 0.8~p.u.\ are failed, and the cascade terminates when no further trips occur or the power flow diverges. The demand not served is:
\begin{equation}
    \text{DNS} = 1 - \frac{\sum_{i \in \mathcal{V}_{\text{served}}} P_{d,i}^{\text{post}}}{\sum_{i \in \mathcal{V}} P_{d,i}^{\text{pre}}}
    \label{eq:dns}
\end{equation}
where $P_{d,i}^{\text{pre}}$ and $P_{d,i}^{\text{post}}$ denote the active load demand at bus $i$ before and after the cascade. Four target labels are extracted per sample: edge labels $\mathbf{y}_{\text{edge}}$, node labels $\mathbf{y}_{\text{node}}$, a binary severity label, and the continuous DNS value. For multi-round training, these are decomposed into per-round targets $\mathbf{y}_{\text{edge}}^{(r)}$ and $\mathbf{y}_{\text{node}}^{(r)}$. The train/validation/test split is performed at the operating-point level to prevent data leakage. Table~\ref{tab:dataset_stats} summarizes the dataset configuration.

\begin{table}[!t]
\centering
\caption{Dataset Configuration for Each IEEE Test System}
\label{tab:dataset_stats}
\renewcommand{\arraystretch}{1.15}
\setlength{\tabcolsep}{5pt}
\begin{tabular}{lcc}
\hline\hline
\textbf{Parameter} & \textbf{IEEE 24-Bus} & \textbf{IEEE 118-Bus} \\
\hline
Buses / branches / directed edges & 24 / 38 / 76 & 118 / 186 / 372 \\
Total samples & 20{,}000 & 20{,}000 \\
Train / Val / Test split & 70\% / 15\% / 15\% & 70\% / 15\% / 15\% \\
Cascade rate & 51.9\% & 16.0\% \\
Edge positive rate & $\sim$2.5\% & $\sim$0.08\% \\
Node positive rate & $\sim$6.7\% & $\sim$7.2\% \\
Severity (Safe / Unsafe) & 71\% / 29\% & 84\% / 16\% \\
Load scaling range & $[0.6, 1.4]$ & $[1.0, 1.6]$ \\
N-$k$ distribution (1/2/3/4) & 70/25/5/-- & 10/45/35/10\% \\
Hard-trip / soft-trip threshold & 120\% / 100\% & 120\% / 95\% \\
Max cascade rounds $R_{\max}$ & 20 & 30 \\
Max observed cascade depth & 5 rounds & 2 rounds \\
\hline\hline
\end{tabular}
\end{table}

\subsection{Input Representation}
\label{subsec:features}
 
The input graph is constructed from the pre-contingency steady-state power flow solution (after the initial outage, before cascade propagation). Tables~\ref{tab:node_features} and~\ref{tab:edge_features} detail the node and edge feature vectors.
 
\begin{table}[!t]
\centering
\caption{Node Feature Vector $\mathbf{x}_i \in \mathbb{R}^{11}$ for Each Bus}
\label{tab:node_features}
\renewcommand{\arraystretch}{1.15}
\setlength{\tabcolsep}{5pt}
\begin{tabular}{clc}
\hline\hline
\textbf{Index} & \textbf{Feature} & \textbf{Source} \\
\hline
0    & Bus type (PQ/PV/Slack)                  & Network topology \\
1--2 & Voltage magnitude, angle                & AC power flow \\
3--4 & Net $P$, $Q$ injection (MW, MVAr)       & AC power flow \\
5--6 & Load $P_d$, $Q_d$ (MW, MVAr)            & Load tables \\
7--8 & Generator $P_g$ (MW), status            & Generator tables \\
9    & Topological degree                      & Network topology \\
10   & Initial contingency mask                & Contingency def. \\
\hline\hline
\end{tabular}
\end{table}
 
All electrical quantities are obtained from the validated Newton-Raphson AC power flow solution. The initial contingency masks encode which components were removed as the trigger event, providing the model with the necessary context to predict the subsequent cascade. In the multi-round setting, these masks are updated at each round to reflect the cumulative failure state.
 
\begin{table}[!t]
\centering
\caption{Edge Feature Vector $\mathbf{e}_{ij} \in \mathbb{R}^{12}$ for Each Directed Edge}
\label{tab:edge_features}
\renewcommand{\arraystretch}{1.15}
\setlength{\tabcolsep}{5pt}
\begin{tabular}{clc}
\hline\hline
\textbf{Index} & \textbf{Feature} & \textbf{Source} \\
\hline
0    & Loading percentage (\% of rating) & AC power flow \\
1--4 & $P_{\text{from}}$, $Q_{\text{from}}$, $P_{\text{to}}$, $Q_{\text{to}}$ (MW, MVAr) & AC power flow \\
5--6 & Resistance $r$, reactance $x$ (p.u.) & Line parameters \\
7    & Thermal rating $I_{\max}$ (kA) & Line parameters \\
8    & In-service status (binary) & Network state \\
9    & Parallel circuits & Line parameters \\
10   & Initial contingency mask & Contingency def. \\
11   & Branch type (line/transformer) & Network topology \\
\hline\hline
\end{tabular}
\end{table}
 
\subsection{PI-GN-JODE Architecture}
\label{subsec:architecture}
 
The PI-GN-JODE architecture consists of four processing stages that mirror the physical processes underlying cascading failures: (1)~an edge-conditioned GNN encoder, (2)~a neural ODE for continuous dynamics, (3)~a jump process handler for discrete relay trips, and (4)~multi-task decoders with Kirchhoff physics regularization. In one-shot mode, these stages execute once. In multi-round mode, they are unrolled across cascade rounds with shared weights, producing a temporal sequence of predictions. The complete one-shot forward pass is:
\begin{equation}
\begin{aligned}
\mathcal{G}(\mathbf{X}, \mathbf{E})
&\xrightarrow{\mathrm{GNN}} \mathbf{H}_0
\xrightarrow{\mathrm{ODE}} \mathbf{H}_c
\xrightarrow{\mathrm{Jump}} \mathbf{H}_f \\
&\xrightarrow{\mathrm{Decoders}}
(\hat{\mathbf{y}}_{\mathrm{edge}},
 \hat{\mathbf{y}}_{\mathrm{node}},
 \hat{\mathbf{y}}_{\mathrm{sev}},
 \hat{y}_{\mathrm{dns}})
\label{eq:forward_pass}
\end{aligned}
\end{equation}

\begin{figure*}[!t]
\centering
\includegraphics[width=\textwidth]{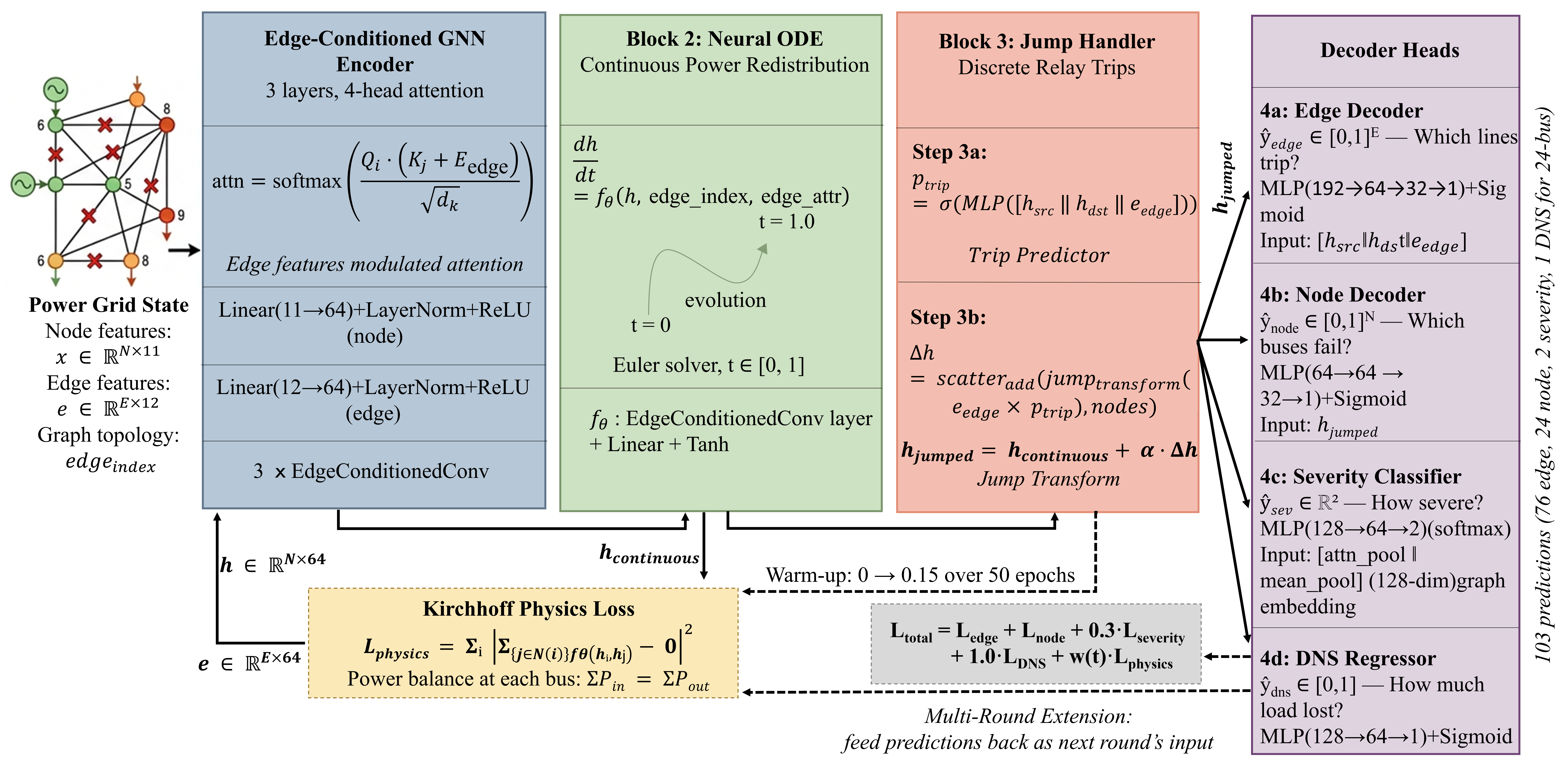}
\caption{PI-GN-JODE architecture. The input graph is processed by a GNN encoder, neural ODE, jump handler, and physics-regularized multi-task decoders to generate four predictions.}
\label{fig:architecture}
\end{figure*}

\subsubsection{Edge-Conditioned GNN Encoder}
 
The encoder transforms raw features into $d$-dimensional hidden representations through $L$ layers of edge-conditioned message passing with multi-head attention. Unlike standard GCNs~\cite{kipf2017semi} and GATs~\cite{velickovic2018graph}, our convolution incorporates branch electrical properties directly into the attention computation following~\cite{simonovsky2017dynamic}.
 
The input features are first projected into the hidden space:
\begin{align}
    \mathbf{h}_i^{(0)} &= \text{LayerNorm}(\text{ReLU}(\mathbf{W}_v \mathbf{x}_i + \mathbf{b}_v)) \label{eq:node_proj} \\
    \mathbf{e}_{ij}' &= \text{LayerNorm}(\text{ReLU}(\mathbf{W}_e \mathbf{e}_{ij} + \mathbf{b}_e)) \label{eq:edge_proj}
\end{align}
where $\mathbf{W}_v \in \mathbb{R}^{d \times d_v}$ and $\mathbf{W}_e \in \mathbb{R}^{d \times d_e}$. At each layer $l$, multi-head attention is computed with edge modulation. For each head $k = 1, \ldots, K$:
\begin{equation}
    \alpha_{ji}^{(k)} = \frac{\exp\!\left(\frac{\mathbf{q}_i^{(k)\top} (\mathbf{k}_j^{(k)} + \mathbf{W}_E^{(k)} \mathbf{e}_{ji}')}{\sqrt{d/K}}\right)}{\sum_{j' \in \mathcal{N}(i)} \exp\!\left(\frac{\mathbf{q}_i^{(k)\top} (\mathbf{k}_{j'}^{(k)} + \mathbf{W}_E^{(k)} \mathbf{e}_{j'i}')}{\sqrt{d/K}}\right)}
    \label{eq:attention}
\end{equation}
where $\mathbf{q}_i^{(k)} = \mathbf{W}_Q^{(k)} \mathbf{h}_i^{(l-1)}$, $\mathbf{k}_j^{(k)} = \mathbf{W}_K^{(k)} \mathbf{h}_j^{(l-1)}$ are the query and key projections, $\mathcal{N}(i)$ denotes the neighbors of node $i$, and $\mathbf{W}_E^{(k)}$ is a per-head edge projection. The edge term $\mathbf{W}_E^{(k)} \mathbf{e}_{ji}'$ modulates the key, allowing branch impedance and loading to directly influence message weights. Each layer also computes an MLP-based message $\mathbf{m}_{ji} = \text{MLP}([\mathbf{h}_j^{(l-1)} \| \mathbf{h}_i^{(l-1)} \| \mathbf{e}_{ji}'])$ and updates the node state with a residual connection:
\begin{equation}
\begin{aligned}
\mathbf{h}_i^{(l)} &=
\operatorname{LayerNorm}\!\Big(
\mathbf{h}_i^{(l-1)} \\
&\quad + \operatorname{Dropout}\!\big(
\operatorname{Attn}_i^{(l)} + \operatorname{FFN}(\mathbf{m}_i^{(l)})
\big)
\Big)
\label{eq:layer_update}
\end{aligned}
\end{equation}
where $\text{FFN}$ is a two-layer feed-forward network with GELU activation~\cite{hendrycks2016gaussian} and layer normalization~\cite{ba2016layer}. After $L$ layers, the encoder produces node embeddings $\mathbf{H}_0 \in \mathbb{R}^{|\mathcal{V}| \times d}$ and edge embeddings $\mathbf{E}' \in \mathbb{R}^{|\mathcal{E}| \times d}$.

\subsubsection{Neural ODE for Continuous Dynamics}
 
Between discrete relay trip events, power flows redistribute continuously through the network as the system seeks a new equilibrium. We model this temporal evolution of node embeddings using a neural ODE~\cite{chen2018neuralode}:
\begin{equation}
    \frac{d\mathbf{H}}{dt} = f_\theta(\mathbf{H}(t), \mathcal{G}), \quad \mathbf{H}(0) = \mathbf{H}_0
    \label{eq:node_ode}
\end{equation}
where $f_\theta$ is parameterized by a single edge-conditioned graph convolution layer followed by a linear transformation with $\tanh$ activation:
\begin{equation}
    f_\theta(\mathbf{H}, \mathcal{G}) = \tanh(\mathbf{W}_f \cdot \text{EdgeCondConv}(\mathbf{H}, \mathcal{G}) + \mathbf{b}_f)
    \label{eq:ode_func}
\end{equation}
The $\tanh$ activation bounds the derivative, preventing unbounded state growth during integration. The ODE is integrated from $t = 0$ to $t = T$ using a fixed-step Euler solver:
\begin{equation}
    \mathbf{H}_c = \mathbf{H}_0 + T \cdot f_\theta(\mathbf{H}_0, \mathcal{G})
    \label{eq:euler_step}
\end{equation}
with $T = 1.0$ treated as an abstract integration horizon rather than physical time. The Euler solver with a single function evaluation is chosen for memory efficiency, as adaptive solvers store activations for each sub-step, causing prohibitive GPU memory consumption on batched graphs.

\subsubsection{Jump Process Handler}
 
Cascading failures are inherently hybrid dynamical systems: continuous power flow redistribution is punctuated by discrete relay trip events. The jump handler models these transitions in the latent space through two stages.
 
First, an MLP-based trip predictor estimates the probability that each edge will trip:
\begin{equation}
    p_{ij}^{\text{trip}} = \sigma\!\left(\text{MLP}_{\text{trip}}([\mathbf{h}_i^c \| \mathbf{h}_j^c \| \mathbf{e}_{ij}'])\right)
    \label{eq:trip_pred}
\end{equation}
where $\mathbf{h}_i^c, \mathbf{h}_j^c \in \mathbf{H}_c$ are the post-ODE node embeddings. For edges with $p_{ij}^{\text{trip}} > 0.5$, a jump transform MLP computes state perturbations at both endpoints:
\begin{equation}
    \boldsymbol{\delta}_s = \text{MLP}_{\text{jump}}([\mathbf{h}_s^c \| \mathbf{e}_{ij}']), \quad s \in \{i, j\}
    \label{eq:jump_transform}
\end{equation}
These perturbations are aggregated across all tripped edges via scatter-add operations. The final node embeddings blend the continuous and jumped states using a learnable scale parameter $\gamma$ (initialized at 0.1):
\begin{equation}
    \mathbf{h}_i^f = (1 - w_i) \cdot \mathbf{h}_i^c + w_i \cdot (\mathbf{h}_i^c + \gamma \cdot \Delta\mathbf{h}_i)
    \label{eq:blend}
\end{equation}
where $w_i = \max_{j:(i,j) \in \mathcal{E}} p_{ij}^{\text{trip}}$ quantifies the maximum trip influence on node $i$. Nodes far from any tripped edge retain their continuous-state embeddings, while nodes adjacent to tripped edges incorporate the discrete perturbation. The implementation is fully vectorized using batched MLP calls and scatter-add aggregation.

\subsubsection{Prediction Heads and Physics Regularization}
\label{subsubsec:decoders}
 
Three decoder heads produce the four prediction targets from the final node embeddings $\mathbf{H}_f$. The node failure decoder maps each node embedding through an MLP ($d \to d \to d/2 \to 1$) to a failure logit. The edge failure decoder concatenates source, destination, and edge embeddings ($3d \to d \to d/2 \to 1$) to produce per-edge trip logits. For graph-level predictions, an attention-based pooling mechanism learns which buses are most critical:
\begin{equation}
\begin{aligned}
\mathbf{g} &= \left[
\sum_{i \in \mathcal{V}} \alpha_i \mathbf{h}_i^f
~\Big\|~
\frac{1}{|\mathcal{V}|}\sum_{i \in \mathcal{V}} \mathbf{h}_i^f
\right] \in \mathbb{R}^{2d}, \\
\alpha_i &= \operatorname{softmax}_i\!\left(
\operatorname{MLP}_{\mathrm{attn}}(\mathbf{h}_i^f)
\right)
\label{eq:graph_embed}
\end{aligned}
\end{equation}
A severity head ($2d \to d \to 2$) produces classification logits and a DNS head ($2d \to d \to 1$) with sigmoid activation outputs the continuous demand-not-served prediction.
 
To encourage physically consistent internal representations, we introduce a soft Kirchhoff power balance regularizer. Two lightweight extractors learn approximate power quantities from the hidden states: a power injection predictor $\hat{\mathbf{p}}_i = \text{MLP}_{\text{inj}}(\mathbf{h}_i^f) \in \mathbb{R}^2$ and a flow predictor $\hat{\mathbf{f}}_{ij} = \text{MLP}_{\text{flow}}([\mathbf{h}_i^f \| \mathbf{h}_j^f]) \in \mathbb{R}^2$, each outputting active and reactive power components. The Kirchhoff violation loss penalizes power imbalance at each node:
\begin{equation}
    \mathcal{L}_{\text{physics}} = \frac{1}{|\mathcal{V}|} \sum_{i \in \mathcal{V}} \left\|\hat{\mathbf{p}}_i + \sum_{j:(j,i) \in \mathcal{E}} \hat{\mathbf{f}}_{ji} - \sum_{j:(i,j) \in \mathcal{E}} \hat{\mathbf{f}}_{ij}\right\|_2^2
    \label{eq:physics_loss}
\end{equation}
This acts as a soft regularizer: the model is not required to solve the power flow equations exactly, but its learned representations are encouraged to respect fundamental conservation laws.

\subsubsection{Multi-Round Autoregressive Extension}
\label{subsubsec:multiround}
 
In the multi-round formulation, the base architecture is applied iteratively with shared weights across cascade rounds $r = 0, 1, \ldots, R-1$:
\begin{equation}
\begin{aligned}
\mathcal{G}^{(r)}
&\xrightarrow{\mathrm{GNN}} \mathbf{H}_0^{(r)}
\xrightarrow{\mathrm{ODE}} \mathbf{H}_c^{(r)}
\xrightarrow{\mathrm{Jump}} \mathbf{H}_f^{(r)} \\
&\xrightarrow{\mathrm{Decoders}}
\left(\hat{\mathbf{y}}_{\mathrm{edge}}^{(r)},
\hat{\mathbf{y}}_{\mathrm{node}}^{(r)}\right)
\label{eq:multiround_forward}
\end{aligned}
\end{equation}
 
The graph state $\mathcal{G}^{(r)}$ is updated between rounds to reflect the cumulative failures. The edge and node contingency masks are augmented with the failures predicted (or observed, during teacher forcing) in all previous rounds, and a learnable \textit{state updater} module transforms the node embeddings to account for the changed topology:
\begin{equation}
    \tilde{\mathbf{h}}_i^{(r)} = \text{MLP}_{\text{update}}\!\left([\mathbf{h}_i^{f,(r-1)} \| \mathbf{c}_i^{(r)}]\right)
    \label{eq:state_update}
\end{equation}
where $\mathbf{c}_i^{(r)} \in \mathbb{R}^2$ encodes the local failure context at node $i$ (whether the node itself failed and the fraction of incident edges that tripped in the previous round). This updated representation is re-encoded through the GNN at the next round, reflecting how the electrical state of the entire network changes when lines trip.
 
Severity and DNS predictions are produced only at the final round, as these are system-level outcomes of the complete cascade. Edge and node failure predictions are produced at every round, providing the temporal trajectory of failure progression. The cascade terminates when the model predicts no further failures (all $p_{ij}^{\text{trip}} < 0.5$) or when the maximum round count $R$ is reached.

\subsection{Training and Evaluation}
\label{subsec:training}
 
\subsubsection{Loss Function and Optimization}
 
The total one-shot training loss is a weighted sum of five components:
\begin{equation}
    \mathcal{L} = \lambda_n \mathcal{L}_{\text{node}} + \lambda_e \mathcal{L}_{\text{edge}} + \lambda_s \mathcal{L}_{\text{sev}} + \lambda_d \mathcal{L}_{\text{dns}} + \lambda_p(t) \mathcal{L}_{\text{physics}}
    \label{eq:total_loss}
\end{equation}
where $\mathcal{L}_{\text{node}}$ and $\mathcal{L}_{\text{edge}}$ are binary cross-entropy with positive-class weights (capped at 30 and 100, respectively), $\mathcal{L}_{\text{sev}}$ is cross-entropy with square-root inverse frequency weights, and $\mathcal{L}_{\text{dns}}$ is MSE. The weights are $\lambda_n{=}1.0$, $\lambda_e{=}1.0$, $\lambda_s{=}0.3$, $\lambda_d{=}1.0$ (see Table~\ref{tab:hyperparameters} for per-system values).
\begin{equation}
    \lambda_p(t) = 
    \begin{cases}
        \lambda_p^{\text{target}} \cdot t / T_w & \text{if } t < T_w \\
        \lambda_p^{\text{target}} & \text{if } t \geq T_w
    \end{cases}
    \label{eq:warmup}
\end{equation}
where $\lambda_p^{\text{target}} = 0.15$ and $T_w$ is the warm-up duration. This schedule prevents the physics loss from destabilizing early training, when hidden representations have not yet learned meaningful power-related features.
 
For multi-round training, the loss is summed over cascade rounds with a discount factor $\beta = 0.95$ that gives slightly less weight to later rounds, where autoregressive error accumulation makes the targets inherently noisier:
\begin{equation}
\begin{aligned}
\mathcal{L}_{\mathrm{MR}}
&= \sum_{r=0}^{R-1} \beta^r
\left(
\lambda_n \mathcal{L}_{\mathrm{node}}^{(r)}
+ \lambda_e \mathcal{L}_{\mathrm{edge}}^{(r)}
\right) \\
&\quad + \lambda_s \mathcal{L}_{\mathrm{sev}}^{(R)}
+ \lambda_d \mathcal{L}_{\mathrm{dns}}^{(R)}
+ \lambda_p(t)\mathcal{L}_{\mathrm{physics}}
\label{eq:multiround_loss}
\end{aligned}
\end{equation}
where the severity and DNS losses are computed only at the final round.
 
Multi-round training employs scheduled sampling~\cite{bengio2015scheduled} to bridge the gap between training (where ground-truth states can be provided) and inference (where the model must rely on its own predictions). The teacher forcing ratio $\tau$ controls the probability of using ground-truth versus predicted failures for state updates between rounds:
\begin{equation}
    \tau(t) = \max\!\left(0, ~1 - \frac{t}{T_{\text{ss}}}\right)
    \label{eq:scheduled_sampling}
\end{equation}
where $T_{\text{ss}}$ is the scheduled sampling decay period. At $\tau = 1.0$ (early training), state updates always use ground-truth failures; at $\tau = 0.0$ (late training and inference), the model is fully autoregressive. This gradual transition allows the model to learn to recover from its own prediction errors.
 
All models are trained using AdamW~\cite{loshchilov2019decoupled} with learning rate $10^{-3}$, weight decay $10^{-5}$, and a OneCycleLR scheduler~\cite{smith2019super} (10\% warm-up, cosine annealing). Gradient norms are clipped to 1.0. Training proceeds for up to 100 epochs with early stopping on validation loss. After training, decision thresholds for node and edge predictions are calibrated on the validation set via grid search to maximize F1. Table~\ref{tab:hyperparameters} summarizes all hyperparameters.
\begin{table}[!t]
\centering
\caption{PI-GN-JODE Hyperparameter Configuration}
\label{tab:hyperparameters}
\renewcommand{\arraystretch}{1.08}
\setlength{\tabcolsep}{2.5pt}
\scriptsize
\begin{tabular}{p{0.36\columnwidth}cc}
\hline\hline
\textbf{Hyperparameter} & \textbf{One-Shot} & \textbf{Multi-Round} \\
\hline
Hidden dimension $d$ & 64 (96$^\dagger$) & 64 (96$^\dagger$) \\
GNN layers $L$ / Attn heads $K$ & 3 / 4 (4 / 4$^\dagger$) & 3 / 4 (4 / 4$^\dagger$) \\
ODE solver / time $T$ & Euler / 1.0 & Euler / 1.0 \\
Dropout & 0.1 & 0.1 \\
Batch size & 16 & 16 \\
Optimizer & \multicolumn{2}{c}{AdamW, lr$=10^{-3}$ ($3{\times}10^{-4}$$^\dagger$), wd$=10^{-5}$} \\
LR scheduler & \multicolumn{2}{c}{OneCycleLR, cosine} \\
Gradient clip norm & 1.0 (0.5$^\dagger$) & 1.0 (0.5$^\dagger$) \\
Max epochs / patience & 100 / 15 & 100 / 30 \\
Weight init & \multicolumn{2}{c}{Xavier uniform~\cite{glorot2010understanding}} \\
$\lambda_n, \lambda_e, \lambda_s, \lambda_d$ & \multicolumn{2}{c}{1.0, 1.0 (0.3$^\dagger$), 0.3, 1.0} \\
$\lambda_p^{\mathrm{target}}$ / $T_w$ & 0.15 / 30 ep. (50$^\dagger$) & 0.15 / 30 ep. (50$^\dagger$) \\
Round discount $\beta$ & -- & 0.95 \\
TF decay $T_{\mathrm{ss}}$ & -- & 60 ep. (24-bus), 120 ep. (118-bus) \\
Max rounds $R$ & -- & 10 (24-bus), 2 (118-bus) \\
\hline\hline
\end{tabular}
\vspace{1mm}
\begin{flushleft}
\footnotesize{Values shown are for the IEEE 24-bus system. $^\dagger$IEEE 118-bus values were different.}
\end{flushleft}
\end{table}
 
\subsubsection{Evaluation Metrics}
 
Given the severe class imbalance (1.4\% edge and 6.9\% node failure rates on the 118-bus system), we use PR-AUC~\cite{saito2015precision} as the primary metric, supplemented by F1 at calibrated thresholds, binary Severity F1, and DNS $R^2$.

\subsubsection{Ablation and Baseline Design}
 
To quantify the contribution of each architectural component, we evaluate four ablation variants that systematically remove individual components while keeping all other settings identical. \textit{GN-JODE} (No Physics) removes Kirchhoff regularization; \textit{PI-GN-ODE} (No Jump) removes the jump handler; \textit{PI-GN-J} (No ODE) removes the neural ODE; and \textit{GNN-Only} removes the ODE, jump handler, and physics loss simultaneously, isolating the combined contribution and interaction effects.
 
Three architecture baselines replace the edge-conditioned encoder: \textit{GCN}~\cite{kipf2017semi} uses standard spectral convolution with only the binary adjacency; \textit{GAT}~\cite{velickovic2018graph} employs multi-head attention without edge conditioning; and \textit{MLP} processes each node and edge independently with no message passing. To validate that the architecture generalizes across simulation frameworks, we additionally evaluate on the PowerGraph benchmark~\cite{varbella2024powergraph}, where the same PI-GN-JODE architecture is trained and tested on data generated using MATPOWER with AC optimal power flow. For multi-round evaluation, we compare the autoregressive model against the one-shot baseline on the same test set to isolate the contribution of temporal unrolling from differences in model capacity.

\section{Results and Discussion}
\label{sec:results}
 
We evaluate PI-GN-JODE across four experimental dimensions: (1)~one-shot cascade prediction with comprehensive ablation analysis on the IEEE~118-bus system, (2)~scalability from 24-bus to 118-bus grids, (3)~architectural benchmarking against the PowerGraph NeurIPS~2024 benchmark, and (4)~multi-round temporal cascade prediction. All results are reported on held-out test sets using group-based splits that prevent operating-point leakage between train, validation, and test partitions.
 
Given the severe class imbalance in cascade prediction, only 1.4\% of edges and 6.9\% of nodes experience failures on the IEEE~118-bus system in any given scenario. We adopt Precision Recall Area Under the Curve (PR-AUC) as the primary metric for edge-level and node-level failure detection, as it is more informative than ROC-AUC under extreme imbalance~\cite{saito2015precision}. We additionally report F1~scores at optimized thresholds, Severity~F1 for the binary severity classification (Safe/Unsafe, with the threshold at DNS $\geq$ 5\%), and the coefficient of determination ($R^2$) for Demand Not Served (DNS) regression. DNS measures the fraction of total system load shed during a cascade, providing a direct estimate of blackout severity.

\subsection{Training Methodology and Convergence Behavior}
\label{sec:results:training}
 
\begin{figure}[!t]
\centering
\includegraphics[width=\columnwidth]{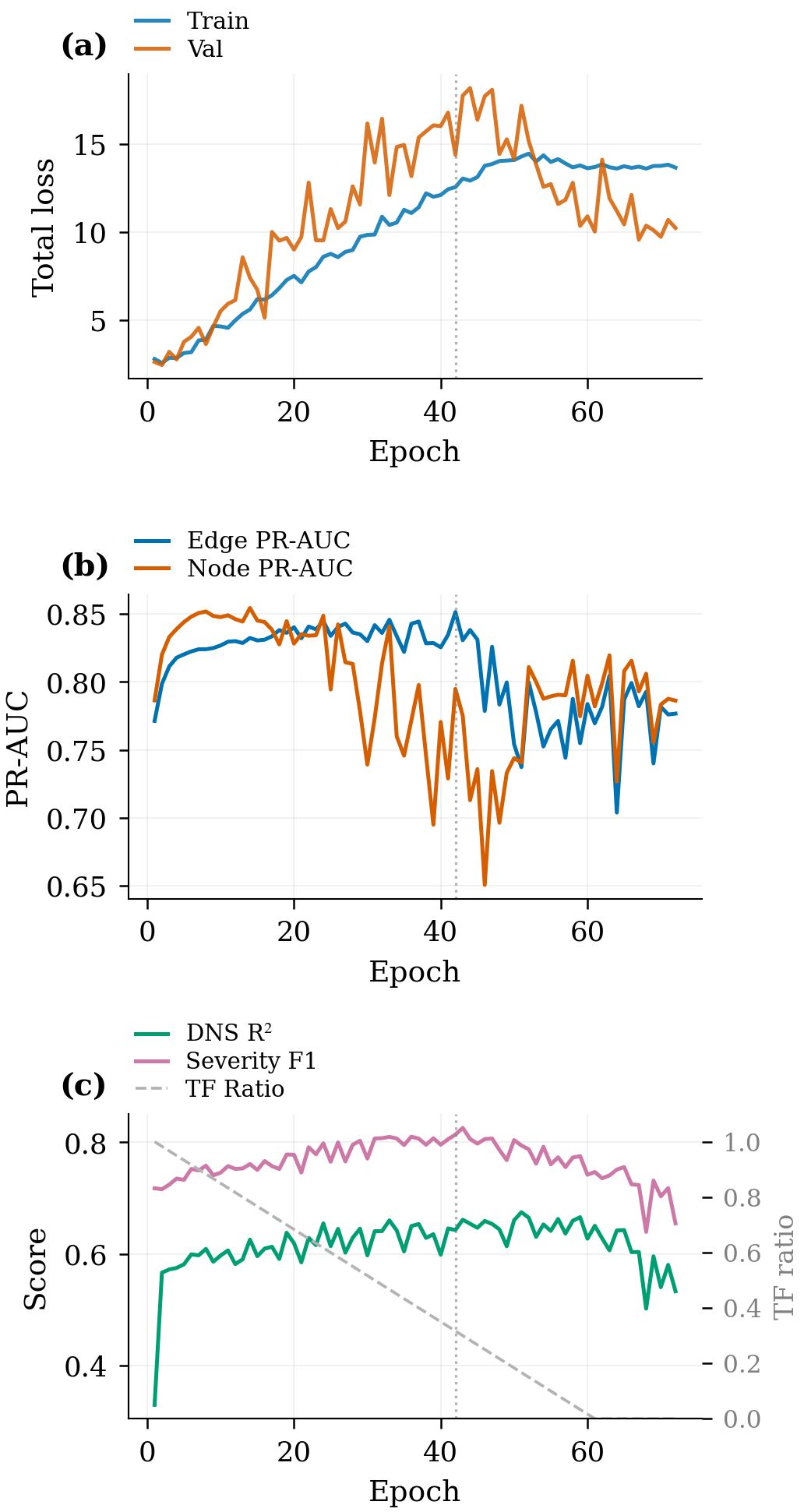}
\caption{Multi-round training on the IEEE~24-bus system over 75 epochs. (a) Loss with best epoch at 42. (b) Edge and node PR-AUC. (c) DNS $R^2$, severity F1, and teacher-forcing ratio.}
\label{fig:train24}
\end{figure}

Multi-round training employs teacher forcing with linear decay from 1.0 to 0.0. Fig.~\ref{fig:train24} shows the 24-bus training dynamics over 75 epochs, with the best validation epoch at 42 (residual TF ratio $\approx 0.4$). The rising training loss reflects TF decay, not divergence. On the 118-bus system, convergence is smoother due to its larger effective dataset (118$\times$372 vs.\ 24$\times$76 features per sample), with the best epoch at 72 (TF $\approx 0.3$) and DNS~$R^2$ exhibiting delayed onset until classification heads partially converge. Across both systems, performance peaks at TF $\approx 0.3$ and degrades below 0.2, suggesting that a residual TF floor may improve autoregressive stability.
 
\subsection{One-Shot Prediction and Component Analysis}
\label{sec:results:oneshot}
 
Table~\ref{tab:118bus_main} presents the one-shot cascade prediction results on the IEEE~118-bus system for the full PI-GN-JODE model, four ablation variants, and three baseline architectures. The full model achieves an Edge PR-AUC of 0.991 and Node PR-AUC of 0.973, demonstrating near-perfect ranking of failure-prone components despite the highly imbalanced label distribution. DNS regression performance is equally strong, with $R^2 = 0.951$ indicating that the model explains over 95\% of the variance in system-level load shedding.
 
\begin{table}[!t]
\centering
\caption{One-Shot Cascade Prediction Results on IEEE 118-Bus System}
\label{tab:118bus_main}
\renewcommand{\arraystretch}{1.15}
\setlength{\tabcolsep}{3.5pt}
\begin{tabular}{lcccccc}
\hline\hline
\textbf{Model} & \textbf{Edge} & \textbf{Edge} & \textbf{Node} & \textbf{Node} & \textbf{Sev.} & \textbf{DNS} \\
 & \textbf{PR-AUC} & \textbf{F1} & \textbf{PR-AUC} & \textbf{F1} & \textbf{F1} & $\boldsymbol{R^2}$ \\
\hline
\multicolumn{7}{l}{\textit{Full model}} \\
PI-GN-JODE       & \textbf{0.991} & \textbf{0.975} & \textbf{0.973} & \textbf{0.929} & \textbf{0.974} & \textbf{0.951} \\
\hline
\multicolumn{7}{l}{\textit{Ablation variants}} \\
No Physics        & 0.960 & 0.929 & 0.932 & 0.898 & 0.923 & 0.858 \\
No Jump           & 0.945 & 0.910 & 0.923 & 0.894 & 0.930 & 0.844 \\
No ODE            & 0.929 & 0.907 & 0.922 & 0.892 & 0.930 & 0.848 \\
GNN-Only          & 0.949 & 0.912 & 0.940 & 0.893 & 0.929 & 0.855 \\
\hline
\multicolumn{7}{l}{\textit{Baseline architectures}} \\
GCN               & 0.948 & 0.911 & 0.925 & 0.890 & 0.949 & 0.912 \\
GAT               & 0.948 & 0.909 & 0.936 & 0.894 & 0.945 & 0.876 \\
MLP               & 0.946 & 0.910 & 0.914 & 0.881 & 0.932 & 0.855 \\
\hline\hline
\end{tabular}
\end{table}
 
Notably, no ablation or baseline matches PI-GN-JODE's joint performance across all six metrics. GCN achieves a competitive DNS~$R^2$ of 0.912, suggesting that simpler architectures suffice for aggregate impact estimation, but its Edge~PR-AUC of 0.948 falls 4.3 points short, confirming that fine-grained failure localization requires PI-GN-JODE's richer representation.

\begin{figure}[!t]
\centering
\includegraphics[width=\columnwidth]{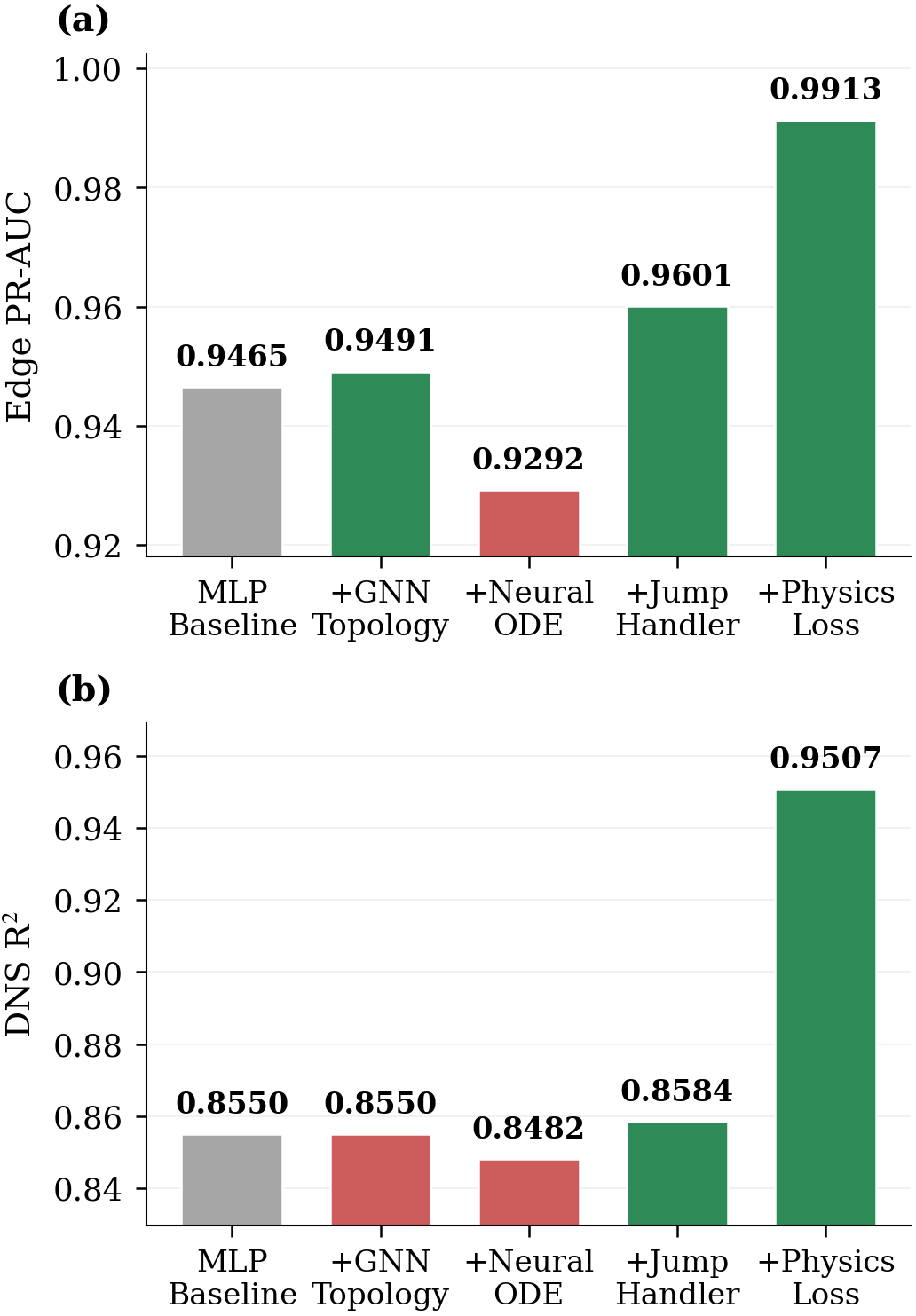}
\caption{Component build-up on IEEE~118-bus. (a) Edge PR-AUC and (b) DNS $R^2$ improve overall, with a brief dip after the Neural ODE that is recovered by the Jump Handler and Physics Loss.}
\label{fig:waterfall}
\end{figure}

To isolate the contribution of each architectural component, we conduct a systematic ablation study where individual modules are removed while all other components and hyperparameters remain fixed. Fig.~\ref{fig:waterfall} presents the cumulative component build-up from the MLP baseline to the full PI-GN-JODE model, illustrating how each addition affects Edge~PR-AUC and DNS~$R^2$.
 
The build-up reveals that the GNN encoder provides modest edge-level gains (+0.003 Edge PR-AUC) but substantial node-level improvement ($0.914 \rightarrow 0.940$), as bus failure prediction depends more on network connectivity than local conditions. The Neural ODE causes a transient dip (Edge PR-AUC: $0.949 \rightarrow 0.929$) due to the harder joint optimization landscape, but the converged model relies heavily on it (removal degrades Edge PR-AUC by 6.2 points). The Jump Handler recovers this dip and adds +3.1 points to Edge PR-AUC. The Physics Loss contributes the largest single improvement to DNS~$R^2$ (+9.2 points, from 0.858 to 0.951), confirming that encoding conservation laws provides a strong inductive bias for system-level impact prediction. On Edge PR-AUC, the Physics Loss and Jump Handler contribute equally (+3.1 points each).
 
Table~\ref{tab:ablation_deltas} summarizes the performance degradation when each component is removed from the full model, providing a complementary view to the build-up analysis.
 
\begin{table}[!t]
\centering
\caption{Ablation Analysis: Performance Degradation ($\Delta$) When Removing Each Component from the Full PI-GN-JODE on IEEE 118-Bus}
\label{tab:ablation_deltas}
\renewcommand{\arraystretch}{1.15}
\setlength{\tabcolsep}{4pt}
\begin{tabular}{lcccc}
\hline\hline
\textbf{Removed} & $\boldsymbol{\Delta}$\textbf{Edge} & $\boldsymbol{\Delta}$\textbf{Node} & $\boldsymbol{\Delta}$\textbf{Sev.} & $\boldsymbol{\Delta}$\textbf{DNS} \\
\textbf{Component} & \textbf{PR-AUC} & \textbf{PR-AUC} & \textbf{F1} & $\boldsymbol{R^2}$ \\
\hline
Physics Loss    & $-$0.031 & $-$0.041 & $-$0.051 & $-$0.093 \\
Jump Handler    & $-$0.046 & $-$0.050 & $-$0.044 & $-$0.107 \\
Neural ODE      & $-$0.062 & $-$0.051 & $-$0.044 & $-$0.103 \\
All (GNN-Only)  & $-$0.042 & $-$0.033 & $-$0.045 & $-$0.096 \\
\hline\hline
\end{tabular}
\end{table}
 
A notable finding is that GNN-Only sometimes yields smaller degradation than individual ablations (e.g., Edge PR-AUC of 0.949 vs No ODE's 0.929). This occurs because GNN-Only converges to a different solution that avoids the ODE's optimization difficulty, while the No ODE variant creates an architectural mismatch by retaining the Jump Handler without the temporally-evolved states it expects. This underscores the synergistic design of PI-GN-JODE's components.
 
\subsection{Scalability and Cross-Simulator Generalization}
\label{sec:results:generalization}
 
\begin{figure}[!t]
\centering
\includegraphics[width=\columnwidth]{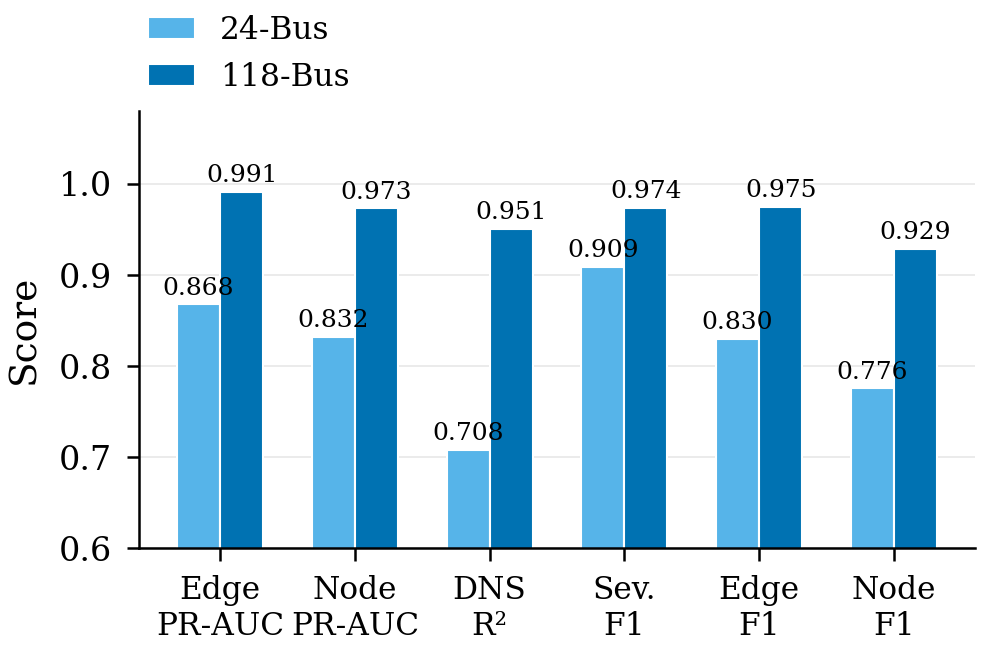}
\caption{One-shot performance comparison between the IEEE~24-bus and IEEE~118-bus systems. Performance improves on all metrics for the larger grid, with the largest gains in DNS $R^2$ and node PR-AUC.}
\label{fig:scaling}
\end{figure}

Fig.~\ref{fig:scaling} compares performance across the IEEE~24-bus and 118-bus systems. Counter-intuitively, the model performs better on the larger grid (Edge PR-AUC: $0.868 \rightarrow 0.991$; DNS $R^2$: $0.708 \rightarrow 0.951$), attributable to more structural context per prediction (5.2 vs 1.9 edge positives per sample), a cleaner severity distribution, and richer message-passing neighborhoods.
 
Table~\ref{tab:24bus_seeds} reports the 24-bus multi-seed results, confirming stable convergence (all standard deviations below 0.02). PI-GN-JODE's advantage over the MLP is more pronounced on the 24-bus system (37.4-point Edge PR-AUC gap vs 4.5 points on 118-bus), reflecting the harder classification task at lower edge positive rates ($\sim$2.5\%).
 
\begin{table}[!t]
\centering
\caption{IEEE 24-Bus One-Shot Results (Mean $\pm$ Std Over 3 Seeds)}
\label{tab:24bus_seeds}
\renewcommand{\arraystretch}{1.15}
\setlength{\tabcolsep}{5pt}
\begin{tabular}{lcc}
\hline\hline
\textbf{Metric} & \textbf{PI-GN-JODE} & \textbf{MLP Baseline} \\
\hline
Edge PR-AUC  & $0.868 \pm 0.009$ & $0.494 \pm 0.012$ \\
Edge F1      & $0.830 \pm 0.009$ & $0.365 \pm 0.015$ \\
Node PR-AUC  & $0.832 \pm 0.008$ & $0.447 \pm 0.014$ \\
Node F1      & $0.776 \pm 0.009$ & $0.389 \pm 0.018$ \\
Severity F1  & $0.909 \pm 0.010$ & $0.703 \pm 0.021$ \\
DNS $R^2$    & $0.708 \pm 0.018$ & $0.412 \pm 0.025$ \\
\hline\hline
\end{tabular}
\end{table}
 
\begin{figure}[!t]
\centering
\includegraphics[width=\columnwidth]{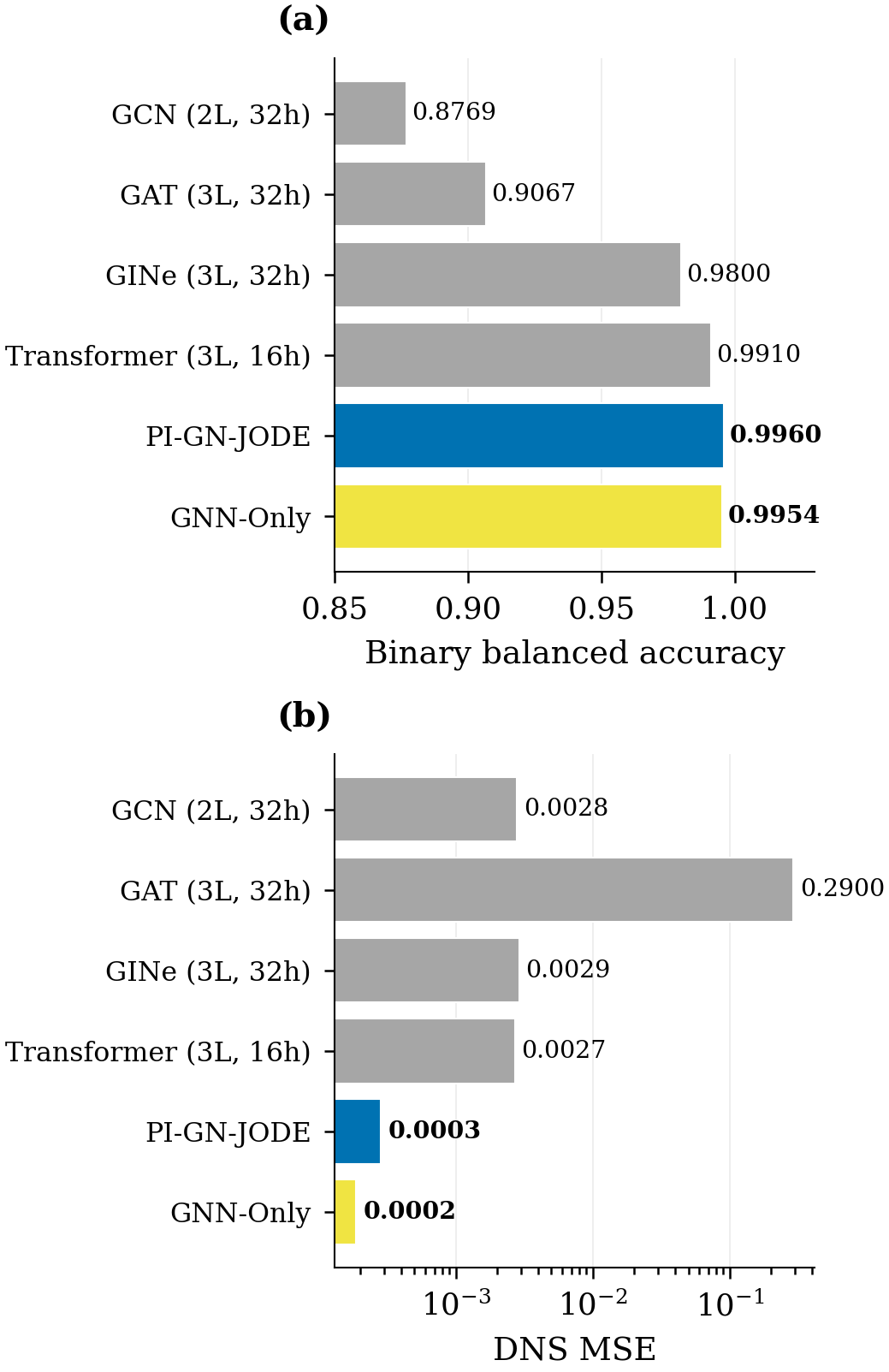}
\caption{Performance comparison on the PowerGraph IEEE~24-bus benchmark~\cite{varbella2024powergraph}. (a) Binary balanced accuracy for cascade detection. (b) DNS mean squared error on a log scale. PI-GN-JODE achieves the highest accuracy with the second-lowest DNS error.}
\label{fig:powergraph}
\end{figure}

To test cross-simulator generalization, we evaluate PI-GN-JODE on the PowerGraph IEEE~24-bus benchmark~\cite{varbella2024powergraph}, training and testing within its data splits for direct comparison. Table~\ref{tab:powergraph} and Fig.~\ref{fig:powergraph} present the results.
 
\begin{table}[!t]
\centering
\caption{Architectural Comparison on the PowerGraph IEEE~24-Bus Benchmark}
\label{tab:powergraph}
\renewcommand{\arraystretch}{1.15}
\setlength{\tabcolsep}{4pt}
\begin{tabular}{llcc}
\hline\hline
\textbf{Source} & \textbf{Model} & \textbf{Bal. Acc.} & \textbf{DNS MSE} \\
\hline
\multicolumn{4}{l}{\textit{Published baselines~\cite{varbella2024powergraph}}} \\
PowerGraph & GCN (2L, 32h)          & 0.8769 & 0.0028 \\
PowerGraph & GAT (3L, 32h)          & 0.9067 & 0.2900 \\
PowerGraph & GINe (3L, 32h)         & 0.9800 & 0.0029 \\
PowerGraph & Transformer (3L, 16h)  & 0.9910 & 0.0027 \\
\hline
\multicolumn{4}{l}{\textit{Ours (PI-GN-JODE architecture)}} \\
Ours & GNN-Only                     & $0.995 \pm 0.001$ & $\mathbf{0.0002}$ \\
Ours & PI-GN-JODE                   & $\mathbf{0.996 \pm 0.002}$ & $0.0003$ \\
\hline\hline
\end{tabular}
\vspace{1mm}
\begin{flushleft}
\footnotesize{All models trained and tested within the PowerGraph framework using identical data splits. PI-GN-JODE uses the same architecture as in our pandapower experiments without modification. DNS MSE reported on logarithmic scale in Fig.~\ref{fig:powergraph}(b). Our GNN-Only achieves the lowest DNS MSE (0.0002); PI-GN-JODE achieves the highest balanced accuracy (0.996).}
\end{flushleft}
\end{table}
 
PI-GN-JODE achieves a balanced accuracy of $0.996 \pm 0.002$, outperforming the best published baseline (Transformer: 0.991) by 0.5 points without any architectural modification, confirming generalization to MATPOWER-based data. On DNS regression, both PI-GN-JODE (MSE = 0.0003) and GNN-Only (MSE = 0.0002) achieve errors an order of magnitude below the best baseline (0.0027); GNN-Only's marginal advantage reflects the single-step setting where ODE and Jump components have limited opportunity to contribute.

\subsection{Temporal Cascade Prediction}
\label{sec:results:multiround}
 
In multi-round mode, PI-GN-JODE autoregressively predicts the cascade state at each round using its own predictions, producing temporal failure trajectories rather than single-step outcome mappings.
 
\begin{figure}[!t]
\centering
\includegraphics[width=\columnwidth]{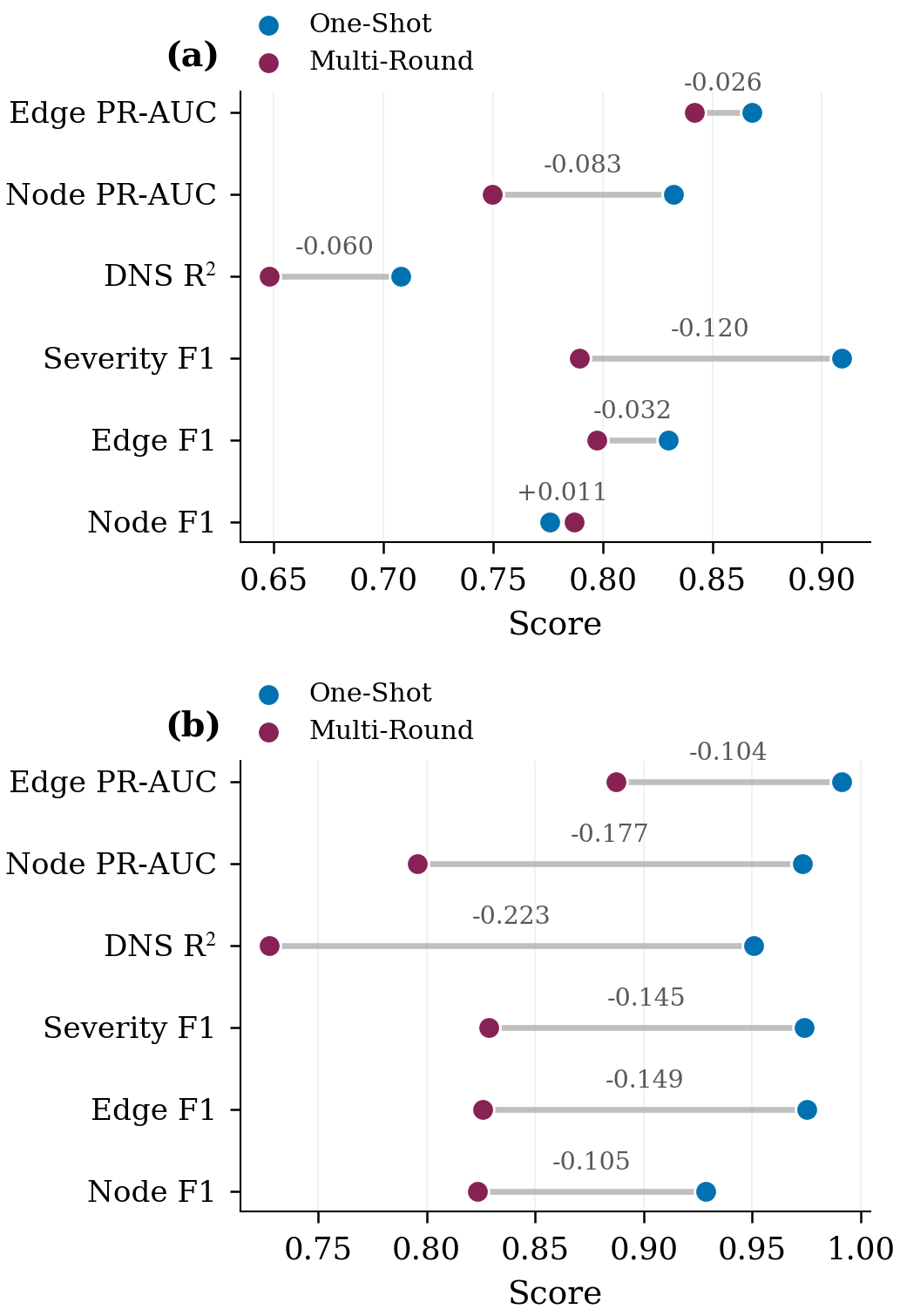}
\caption{One-shot versus multi-round performance comparison. (a) IEEE 24-bus: moderate drops across metrics, with the largest decline in DNS $R^2$. (b) IEEE 118-bus: larger drops across all metrics, again led by DNS $R^2$.}
\label{fig:dumbbell}
\end{figure}

Fig.~\ref{fig:dumbbell} compares one-shot and multi-round performance on both test systems. Table~\ref{tab:multiround} provides the corresponding numerical results.
 
\begin{table}[!t]
\centering
\caption{Multi-Round Cascade Prediction Results}
\label{tab:multiround}
\renewcommand{\arraystretch}{1.15}
\setlength{\tabcolsep}{3pt}
\begin{tabular}{llcccccc}
\hline\hline
\textbf{Grid} & \textbf{Mode} & \textbf{Edge} & \textbf{Node} & \textbf{DNS} & \textbf{Sev.} & \textbf{Edge} & \textbf{Node} \\
 &  & \textbf{PR-AUC} & \textbf{PR-AUC} & $\boldsymbol{R^2}$ & \textbf{F1} & \textbf{F1} & \textbf{F1} \\
\hline
\multirow{2}{*}{24-Bus} & OS  & 0.868 & 0.832 & 0.708 & 0.909 & 0.830 & 0.776 \\
                         & MR  & 0.842 & 0.750 & 0.648 & 0.789 & 0.798 & 0.787 \\
\cline{2-8}
                         & $\Delta$ & $-$0.026 & $-$0.083 & $-$0.060 & $-$0.120 & $-$0.032 & $+$0.011 \\
\hline
\multirow{2}{*}{118-Bus} & OS & 0.991 & 0.973 & 0.951 & 0.974 & 0.975 & 0.929 \\
                          & MR & 0.887 & 0.796 & 0.728 & 0.829 & 0.826 & 0.824 \\
\cline{2-8}
                          & $\Delta$ & $-$0.104 & $-$0.177 & $-$0.223 & $-$0.145 & $-$0.149 & $-$0.105 \\
\hline\hline
\end{tabular}
\vspace{1mm}
\begin{flushleft}
\footnotesize{OS = One-Shot; MR = Multi-Round. $\Delta$ = MR $-$ OS. The 24-bus model predicts up to 5 cascade rounds; the 118-bus model predicts up to 2 rounds (reflecting the network's meshed topology).}
\end{flushleft}
\end{table}
 
On the 24-bus system, the multi-round model exhibits moderate degradation: Edge~PR-AUC decreases by 2.6 points (from 0.868 to 0.842) while DNS~$R^2$ drops by 6.0 points (from 0.708 to 0.648). Notably, Node~F1 \textit{improves} by 1.1 points in the multi-round setting ($0.776 \rightarrow 0.787$), suggesting that the temporal unrolling provides additional signal for bus-level failure detection even as it introduces autoregressive error accumulation on other metrics.
 
On the 118-bus system, gaps are larger (Edge PR-AUC: $-$10.4; DNS $R^2$: $-$22.3) due to error compounding over 372 edges, limited cascade depth (max 2 rounds), and the exceptionally strong one-shot baseline leaving less headroom. Despite this, the multi-round model achieves practically useful performance (Edge PR-AUC 0.887, Node PR-AUC 0.796) while providing operators with failure \textit{trajectories} rather than static snapshots.
 
\subsection{Discussion}
\label{sec:results:discussion}
 
\subsubsection{Architectural Synergy and the Role of Physics}
 
The ablation results confirm that PI-GN-JODE's four components are synergistic: the Neural ODE's transient dip during build-up (Fig.~\ref{fig:waterfall}) and GNN-Only's occasionally smaller degradation than partial ablations (Table~\ref{tab:ablation_deltas}) both indicate an integrated system rather than independently additive modules. The Physics Loss provides the largest single contribution to DNS~$R^2$ (+9.2 points), with its warm-up schedule avoiding early optimization conflicts and enabling 5$\times$ longer productive training.

\subsubsection{Cascade Dynamics and Practical Implications}
 
The maximum cascade depth (3--5 rounds on 24-bus vs 2 rounds on 118-bus) reflects the meshed 118-bus topology, which either absorbs contingencies or collapses in a single wave. PI-GN-JODE produces predictions in $\sim$5\,ms on a GPU ($100$--$400\times$ faster than pandapower), enabling real-time contingency screening with temporal failure trajectories unavailable from static analysis.

\subsubsection{Limitations}
 
Several limitations should be noted. First, the multi-round model exhibits substantial DNS degradation on the 118-bus system ($R^2$ from 0.951 to 0.728), indicating that autoregressive error accumulation remains a challenge for system-level impact estimation on large grids. Second, the cascade simulation uses a simplified relay model (deterministic thermal overload trips) without modeling protection system hidden failures, operator actions, or remedial action schemes that influence real cascades. Third, while we demonstrate scalability from 24 to 118 buses, the transition to realistic bulk power systems with thousands of buses will likely require architectural modifications such as hierarchical graph representations or graph coarsening to manage computational complexity. Finally, the PowerGraph benchmark comparison, while demonstrating that the architecture performs well on data from a different simulator, is limited to the 24-bus topology; extending this validation to larger benchmark grids as they become available will strengthen the generalization claims.

\section{Conclusion}
\label{sec:conclusion}
 
This paper introduced PI-GN-JODE, a unified framework that combines edge-conditioned graph neural networks, Neural ODE continuous dynamics, discrete jump processes, and Kirchhoff-based physics regularization for cascading failure prediction in power grids. Evaluation on the IEEE 24-bus and 118-bus systems demonstrated strong multi-task performance (Edge PR-AUC 0.991, Node PR-AUC 0.973, DNS $R^2$ 0.951 on 118-bus), with ablation studies confirming synergistic component interactions and the physics loss providing the largest single contribution (+9.2 points in $R^2$). The architecture generalizes across simulation frameworks, achieving the highest balanced accuracy (0.996) on the PowerGraph benchmark. The multi-round autoregressive extension provides temporal failure trajectories, though autoregressive error accumulation on larger grids remains an open challenge. Future work includes error-aware autoregressive training, scaling to bulk power systems via hierarchical graph representations, incorporating more realistic cascade dynamics, and real-time contingency screening deployment.

\ifCLASSOPTIONcaptionsoff
  \newpage
\fi
 
\section*{Acknowledgment}
This work was supported in part by the U.S. National Science Foundation (NSF) under Award No. 2509993 and in part by the University of Michigan-Dearborn Experience+ Student Independent Research Grant.
 
\bibliographystyle{IEEEtran}
\bibliography{references}
 
\end{document}